\documentclass[12pt,prd,tightenlines,nofootinbib]{revtex4}
\usepackage{bm}
\usepackage{graphics}
\usepackage{rotating}
\usepackage{epsfig}
\begin{document}
\title{
Semileptonic $\Xi_c$ baryon decays in the relativistic quark model}  
\author{R. N. Faustov}
\author{V. O. Galkin}
\affiliation{Institute of Cybernetics and  Informatics in Education, FRC CSC RAS,
  Vavilov Street 40, 119333 Moscow, Russia}

\begin{abstract}
The form factors of the weak $\Xi_c\to \Xi(\Lambda)$ transitions are
calculated in the framework of the relativistic quark model based on
the quasipotential approach. All relativistic effects including
transformations of the baryon wave functions from the rest to moving
reference frame and
contributions of the intermediate negative energy states are
systematically taken into account. The explicit analytic
expressions which reliably approximate  the momentum transfer $q^2$
dependence of the form factors in the whole accessible kinematical range are given. The calculated form factors are
applied for the evaluation of the semileptonic $\Xi_c\to \Xi \ell\nu_\ell$ and
$\Xi_c\to \Lambda \ell\nu_\ell$ ($\ell=e,\mu$) decay rates,
different asymmetry and polarization parameters within helicity
formalism. The obtained results are compared with available
experimental data and previous calculations.    
\end{abstract}
\vspace*{0.2cm}

\maketitle

\section{Introduction}

This year significant experimental progress has been achieved in
studying weak decays of the charmed $\Xi_c$ baryons. Until now the
absolute branching fractions of both neutral $\Xi_c^0$ and charged
$\Xi_c^+$ baryons were not measured. All decay modes were only measured
relative to  $\Xi_c^0\to\Xi^-\pi^+$ and $\Xi_c^+\to\Xi^-\pi^+\pi^+$
modes \cite{pdg}. This fact significantly complicated comparison of
theoretical predictions with experimental data. However, recently the Belle
Collaboration presented the first measurement of absolute branching
fractions of the neutral $\Xi_c^0$ baryon in three decay modes including
$\Xi_c^0\to\Xi^-\pi^+$ one \cite{belle1}. Its branching
fraction is $Br(\Xi_c^0\to\Xi^-\pi^+)=(1.80\pm0.50\pm0.14)\%$. Then the
absolute branching fractions of its charged partner $\Xi_c$ were
also reported  \cite{belle2}  for
three decay modes including $\Xi_c^+\to\Xi^-\pi^+\pi^+$  with
$Br(\Xi_c^+\to\Xi^-\pi^+\pi^+)=(2.86\pm1.21\pm0.38)\%$. These results
can be combined with $\Xi_c$ branching fractions measured relative to corresponding
modes to get other absolute $\Xi_c$ branching fractions. Thus the
experimental values of $\Xi_c$ semileptonic branching fractions can be
determined.

In this paper we calculate the weak $\Xi_c\to \Xi(\Lambda)$ transition form
factors in the framework of the relativistic quark model based on the
quasipotential approach and use them to evaluate  the
semileptonic branching fractions of the $\Xi_c$ baryon. This model was
successfully applied for studying semileptonic decays of bottom
$\Lambda_b$ \cite{Lambdabsl} and $\Xi_b$ \cite{xib} and charmed
$\Lambda_c$ \cite{Lambdacsl} baryons. The form factors are expressed
as the overlap integrals of the baryon wave functions. The important
advantage of the employed model is the comprehensive inclusion of the
relativistic effects which allows us to explicitly determine the $q^2$
dependence of the form factors in the whole kinematical range, thus
increasing reliability of the results. The calculated form factors can
be then used for the determination of branching fractions and other
important observables which can be measured experimentally. The
results can be confronted with previous theoretical predictions and
new experimental data.

\section{  Form factors of the weak $\Xi_c$ baryon decays}

We consider the matrix element of the weak current $J_\mu^W=V_\mu-A_\mu=\bar
q\gamma_\mu(1-\gamma_5)c$ ($q=s,d$) between $\Xi_c$ and $\Xi$ or
$\Lambda$ baryon states, respectively. In the quasipotential approach
it is expressed
through the overlap integral of the baryon wave functions $\Psi_{B{\bf
    p}_B}$ ($B=\Xi_c,\Xi,\Lambda$) projected on the positive energy
states (${\bf p}_B$ is the baryon momentum) \cite{Lambdabsl}
\begin{equation}\label{mel} 
\langle \Xi(\Lambda)(p_{{\Xi(\Lambda)}}) \vert J^W_\mu \vert \Xi_c(p_{\Xi_c})\rangle
=\int \frac{d^3p\, d^3q}{(2\pi )^6} \bar \Psi_{\Xi(\Lambda)\,{\bf p}_{{\Xi(\Lambda)}}}({\bf
p})\Gamma _\mu ({\bf p},{\bf q})\Psi_{\Xi_c\,{\bf p}_{\Xi_c}}({\bf q}).
\end{equation}
Here $\Gamma _\mu ({\bf p},{\bf
q})$ is the two-particle vertex function which receives relativistic
contributions both from the impulse approximation diagram and from the
diagrams with the intermediate negative-energy states \cite{Lambdabsl}. It is
convenient to consider semileptonic decay in the rest frame of the
initial $\Xi_c$ baryon. Then it is necessary to take  the
final $\Xi(\Lambda)$ in the moving frame \cite{Lambdabsl}.  

The weak $\Xi_c\to \Xi(\Lambda)$ transition matrix elements are parametrized  in terms of
six invariant form factors \cite{giklsh}
\begin{eqnarray}
  \label{eq:ff}
  \langle \Xi(\Lambda)(p_{{\Xi(\Lambda)}},s')|V^\mu|\Xi_c(p_{\Xi_c},s)\rangle&= &\bar
  u_{\Xi(\Lambda)}(p_{{\Xi(\Lambda)}},s')\Bigl[f_1^V(q^2)\gamma^\mu\cr&&\qquad\quad-f_2^V(q^2)i\sigma^{\mu\nu}\frac{q_\nu}{M_{\Xi_c}}+f_3^V(q^2)\frac{q^\mu}{M_{\Xi_c}}\Bigr]
u_{\Xi_c}(p_{\Xi_c},s),\cr
 \langle \Xi(\Lambda)(p_{{\Xi(\Lambda)}},s')|A^\mu|\Xi_c(p,s)\rangle&=& \bar
  u_{\Xi(\Lambda)}(p_{{\Xi(\Lambda)}},s')\Bigl[f_1^A(q^2)\gamma^\mu\cr&&\qquad\quad-f_2^A(q^2)i\sigma^{\mu\nu}\frac{q_\nu}{M_{\Xi_c}}+f_3^A(q^2)\frac{q^\mu}{M_{\Xi_c}}\Bigl]
\gamma_5 u_{\Xi_c}(p_{\Xi_c},s),\qquad
\end{eqnarray}
where $M_{B}$ and  $u_B(p,s)$ are masses and Dirac spinors of
the initial and final baryons ($B=\Xi_c,\Xi,\Lambda$), $q=p_{{\Xi(\Lambda)}}-p_{\Xi_c}$.

Comparing (\ref {mel}) with (\ref{eq:ff}) we obtain these form factors
in the framework of the relativistic quark model. They are expressed through the overlap integrals of
the baryon wave functions \cite{Lambdabsl} which are known from the
calculations of the baryon mass spectra \cite{hbarregge,sbar}.
The explicit expressions are given in Ref.~\cite{Lambdabsl}. They
systematically take into account all relativistic effects including transformation
of baryon wave functions from the rest to moving reference frame and
contributions of the intermediate negative energy states. 

\begin{figure}[hbt]
\centering
  \includegraphics[width=8cm]{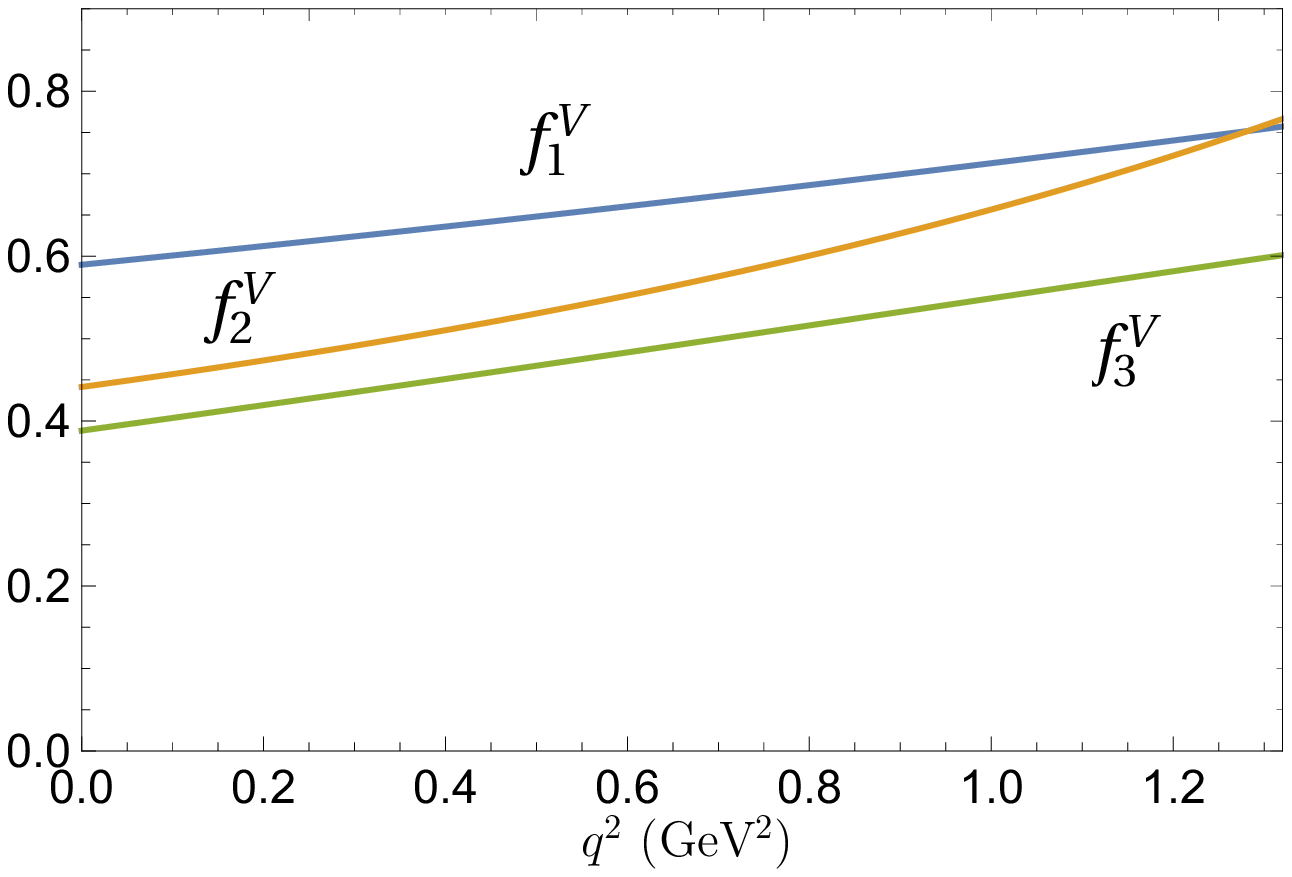}\ \
 \ \includegraphics[width=8cm]{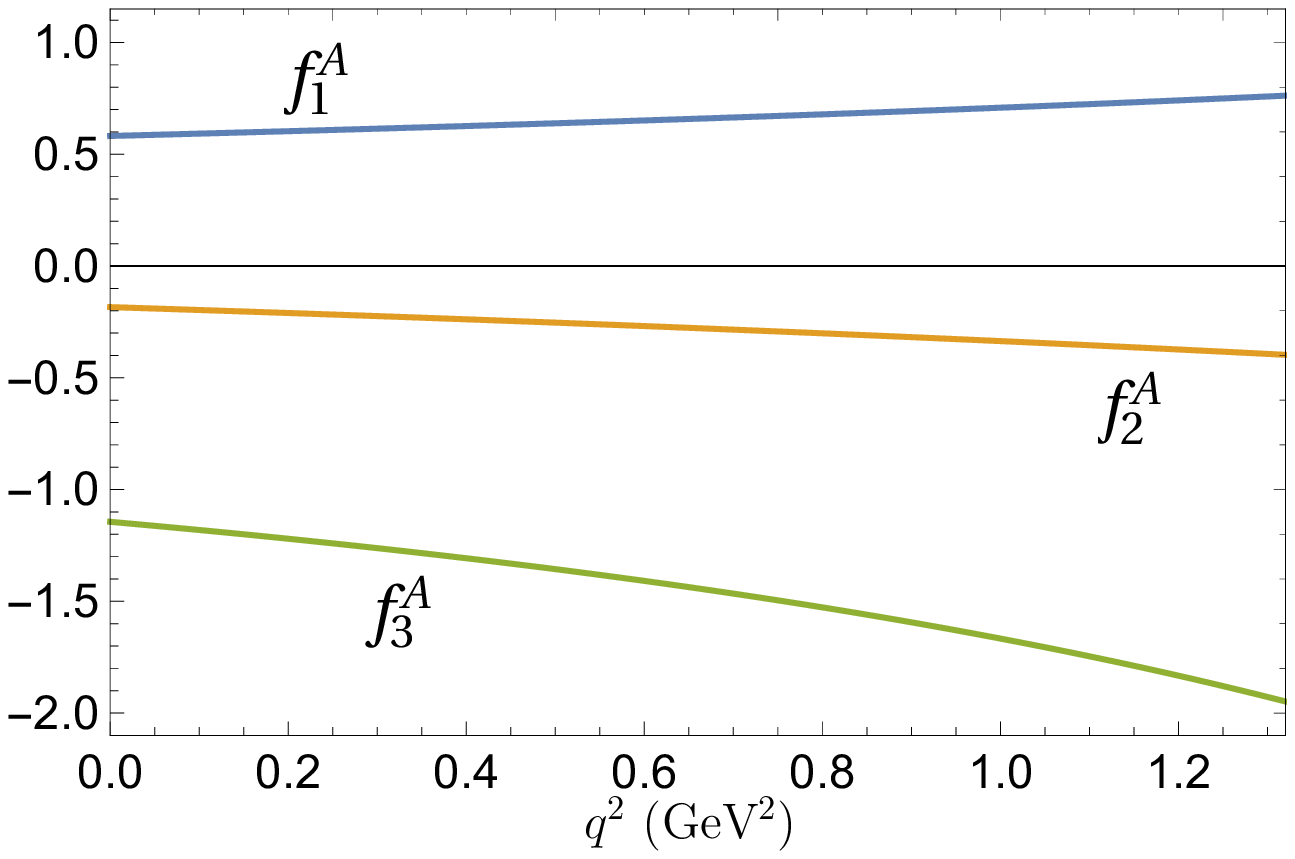}\\
\caption{Form factors of the weak $\Xi_c\to \Xi$ transitions.    } 
\label{fig:ffXicXi}
\end{figure}

\begin{figure}[hbt]
\centering
  \includegraphics[width=8cm]{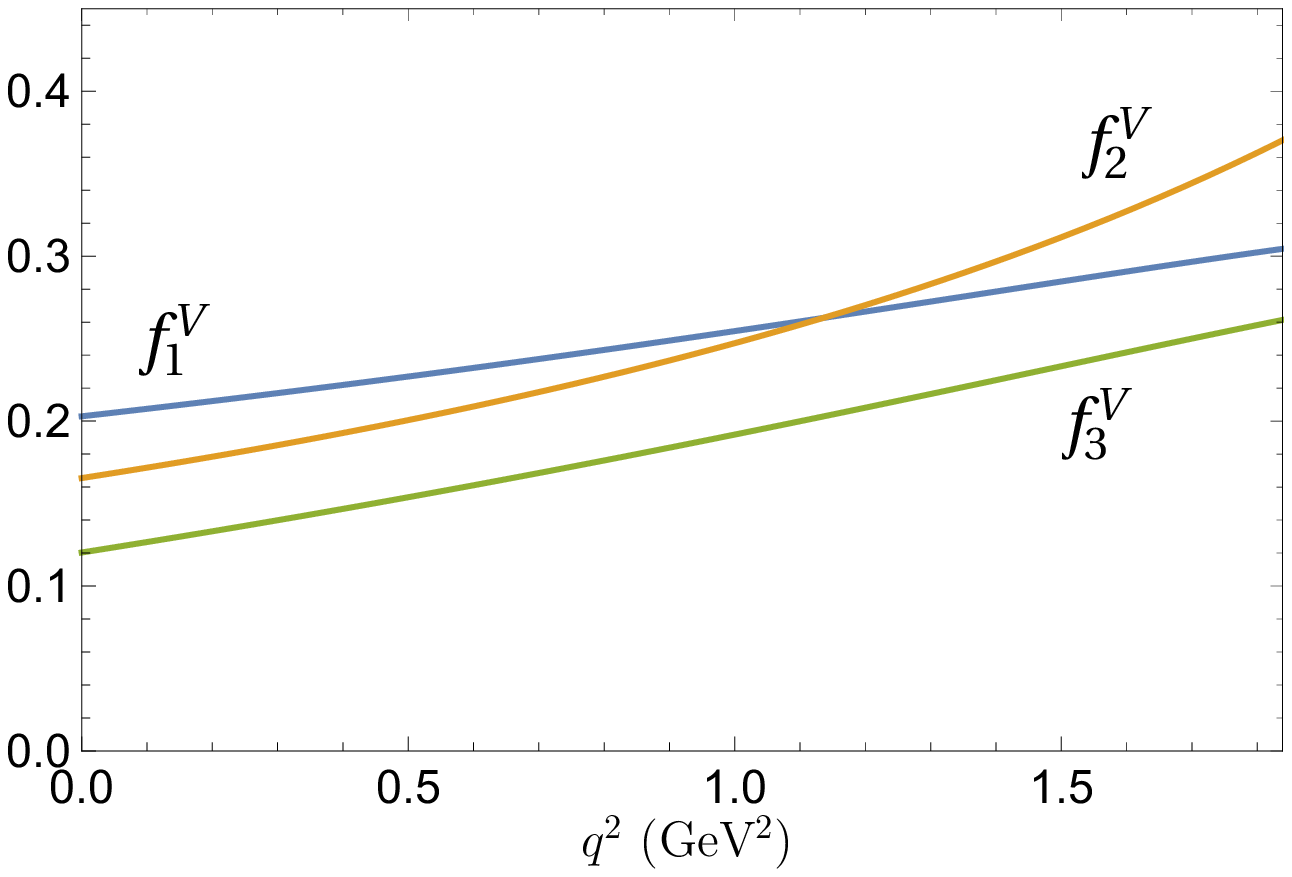}\ \
 \ \includegraphics[width=8cm]{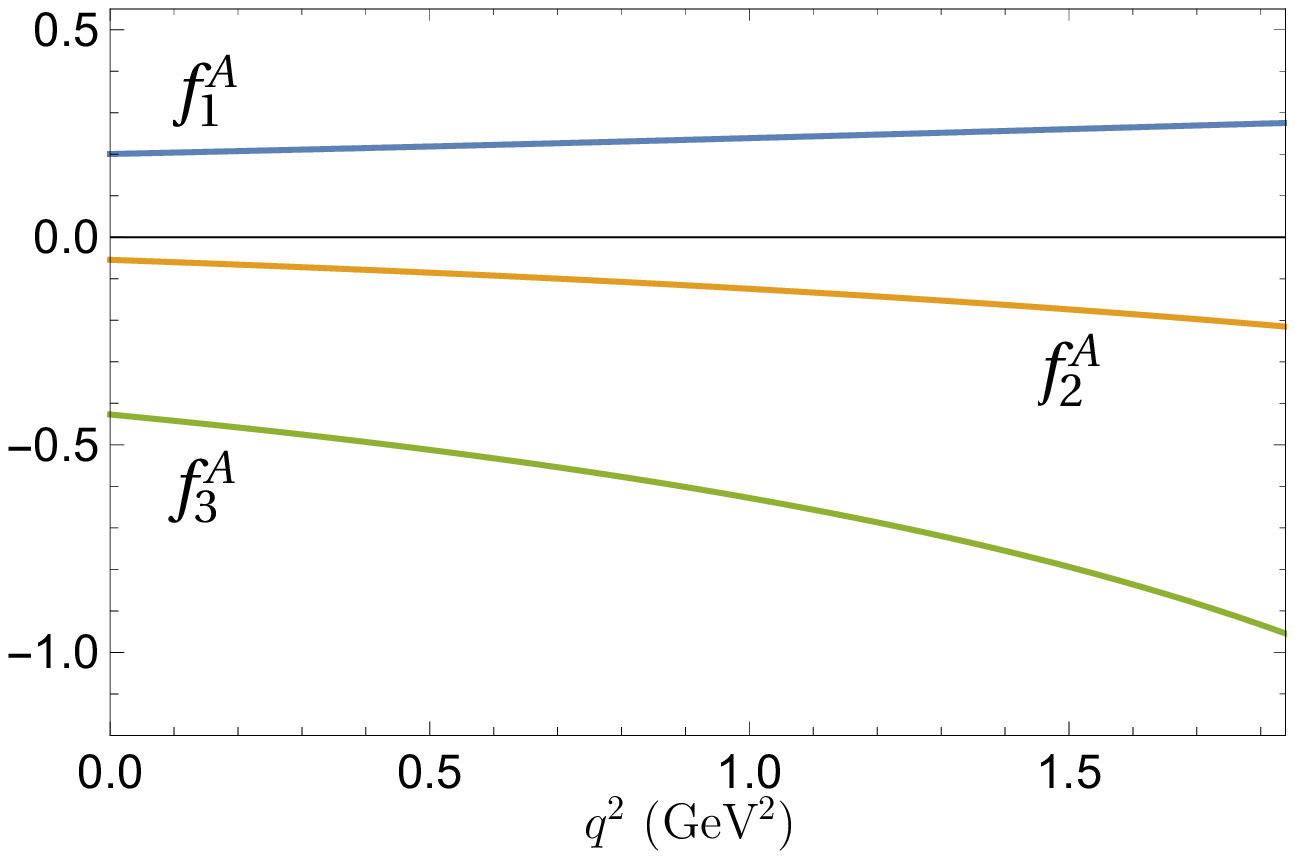}\\
\caption{Form factors of the weak $\Xi_c\to \Lambda$ transitions.    } 
\label{fig:ffXicLambda}
\end{figure}

Following Ref.~\cite{xib} we fit the numerically calculated form
factors by the following analytic expression 
\begin{equation}
  \label{fitff}
  f(q^2)= \frac{1}{{1-q^2/{M_{\rm pole}^2}}} \left\{ a_0 + a_1 z(q^2) +
    a_2 [z(q^2)]^2 \right\},
\end{equation}
where the variable 
\begin{equation}
z(q^2) = \frac{\sqrt{t_+-q^2}-\sqrt{t_+-t_0}}{\sqrt{t_+-q^2}+\sqrt{t_+-t_0}},
\end{equation}
here $t_+=(M_{D_s(D)}+M_{K(\pi)})^2$ and $t_0 = q^2_{\rm max} = (M_{\Xi_c} - M_{\Xi(\Lambda)})^2$.  The pole
masses have the following  values.
For $\Xi_c\to\Xi$ transitions:
$M_{\rm
  pole}\equiv M_{D_s^*}=2.112$ GeV for $f_{1,2}^V$; $M_{\rm
  pole}\equiv M_{D_{s1}}=2.535$ GeV for $f_{1,2}^A$ and $f_{3}^V$; $M_{\rm
  pole}\equiv M_{D_s}=1.969$ GeV for $f_{3}^A$.
For $\Xi_c\to\Lambda$ transitions: $M_{\rm
  pole}\equiv M_{D^*}=2.010$ GeV for $f_{1,2}^V$; $M_{\rm
  pole}\equiv M_{D_{1}}=2.423$ GeV for $f_{1,2}^A$ and $f_{3}^V$;  $M_{\rm
  pole}\equiv M_{D}=1.870$ GeV for $f_{3}^A$.
The fitted values of the parameters $a_0$, $a_1$, $a_2$ as well as the
values of form factors at maximum $q^2=0$ and zero recoil $q^2=q^2_{\rm
  max}$ of the final baryon are given in Tables~\ref{ffXicXi}, \ref{ffXicLambda}. The difference of the fitted
form factors from the calculated ones does not exceed 0.5\%. The form factors
are plotted in Figs.~\ref{fig:ffXicXi}, \ref{fig:ffXicLambda}. We roughly estimate the total
uncertainty of our form factor calculation to be about 5\%.

\begin{table}
\caption{Form factors of the weak $\Xi_c\to \Xi$ transitions. }
\label{ffXicXi}
\begin{ruledtabular}
\begin{tabular}{ccccccc}
& $f^V_1(q^2)$ & $f^V_2(q^2)$& $f^V_3(q^2)$& $f^A_1(q^2)$ & $f^A_2(q^2)$ &$f^A_3(q^2)$\\
\hline
$f(0)$          &0.590 &$0.441$ & $0.388$ & 0.582 & $-0.184$&$-1.144$\\
$f(q^2_{\rm max})$&0.757  &$0.766$ & $0.601$ & 0.762& $-0.397$& $-1.948$ \\
$a_0$      &$0.533$&$0.540$& $0.478$& $0.606$ &$-0.316$&  $-1.285$\\
$a_1$      &$1.323$&$-1.941$&$-0.660$&$-0.482$&$1.894$&$3.174$\\
$a_2$      &$-6.53$&$5.54$& $-12.96$& $1.55$ &$4.19$&  $-14.42$\\
\end{tabular}
\end{ruledtabular}
\end{table}

\begin{table}
\caption{Form factors of the weak $\Xi_c\to \Lambda$ transitions. }
\label{ffXicLambda}
\begin{ruledtabular}
\begin{tabular}{ccccccc}
& $f^V_1(q^2)$ & $f^V_2(q^2)$& $f^V_3(q^2)$& $f^A_1(q^2)$ & $f^A_2(q^2)$ &$f^A_3(q^2)$\\
\hline
$f(0)$          &0.203 &$0.165$ & $0.120$ & 0.201 & $-0.054$&$-0.427$\\
$f(q^2_{\rm max})$&0.304  &$0.370$ & $0.261$ & 0.275& $-0.215$& $-0.954$ \\
$a_0$      &$0.166$&$0.202$& $0.179$& $0.189$ &$-0.148$&  $-0.453$\\
$a_1$      &$0.402$&$-0.142$&$-0.091$&$0.154$&$0.481$&$-0.084$\\
$a_2$      &$-1.04$&$-0.653$& $-2.00$& $-5.15$ &$0.917$&  $1.69$\\
\end{tabular}
\end{ruledtabular}
\end{table}

In Table~\ref{compff} we compare our results for the values of form
factors at $q^2=0$ with previous calculations. Results  \cite{zhao}
are based on the light-front
approach.~\footnote{In Table~\ref{compff} we present the
  ``physical transition form factors'' from Ref.~\cite{zhao} which
  include overlapping factors.} In this paper form factors $f_3^{V,A}(0)$ were
not evaluated. We find good agreement for the form factors
parametrizing both the $\Xi_c\to\Xi$ and $\Xi_c\to\Lambda$ weak transitions. In Ref.~\cite{ass} the light cone QCD sum rules were
employed for the calculation of the $\Xi_c\to \Xi$ transition form
factors. Their form factors $f^V_{1,2,3}(0)$ parametrizing the vector
current significantly differ from our results. The central values \cite{ass} of
the form factor $f^V_{1}(0)$ is about 3 times smaller while
those for $f^V_{2}(0)$ are a factor of about 2 larger
than other predictions and form factors $f^V_{3}(0)$,  $f^A_{2}(0)$ have even opposite signs.     

\begin{table}
  \caption{Comparison of the theoretical predictions for form
    factors of the $\Xi_c$ baryon weak transitions at $q^2=0$.}
  \label{compff}
  \begin{ruledtabular}
    \begin{tabular}{ccccccc}
& $f_1^V(0)$ & $f_2^V(0)$ & $f_3^V(0)$& $f_1^A(0)$ & $f_2^A(0)$ &
$f_3^A(0)$\\
\hline
$\Xi_c\to\Xi$\\
present work& 0.590&0.441& 0.388& 0.582& $-0.184$ & $-1.144$\\
      \cite{zhao}& 0.567& 0.305&    &0.491 & 0.046 \\
\cite{ass}&$0.194\pm0.050$&$0.880\pm0.227$&$-1.14\pm0.30$&$0.311\pm0.081$ &$0.373\pm0.094$&$-0.771\pm0.200$\\      
$\Xi_c\to\Lambda$\\
present work& 0.203&0.165& 0.120& 0.201& $-0.054$ & $-0.427$\\
\cite{zhao}& 0.253& 0.149&    &0.217 & 0.019 \\
\end{tabular}
      \end{ruledtabular}
 \end{table}

\section{Semileptonic $\Xi_c\to\Xi\ell\nu_\ell$ and 
$\Xi_c\to \Lambda\ell\nu_\ell$ decays}

\begin{figure}[hbt]
  \centering
 \includegraphics[width=8cm]{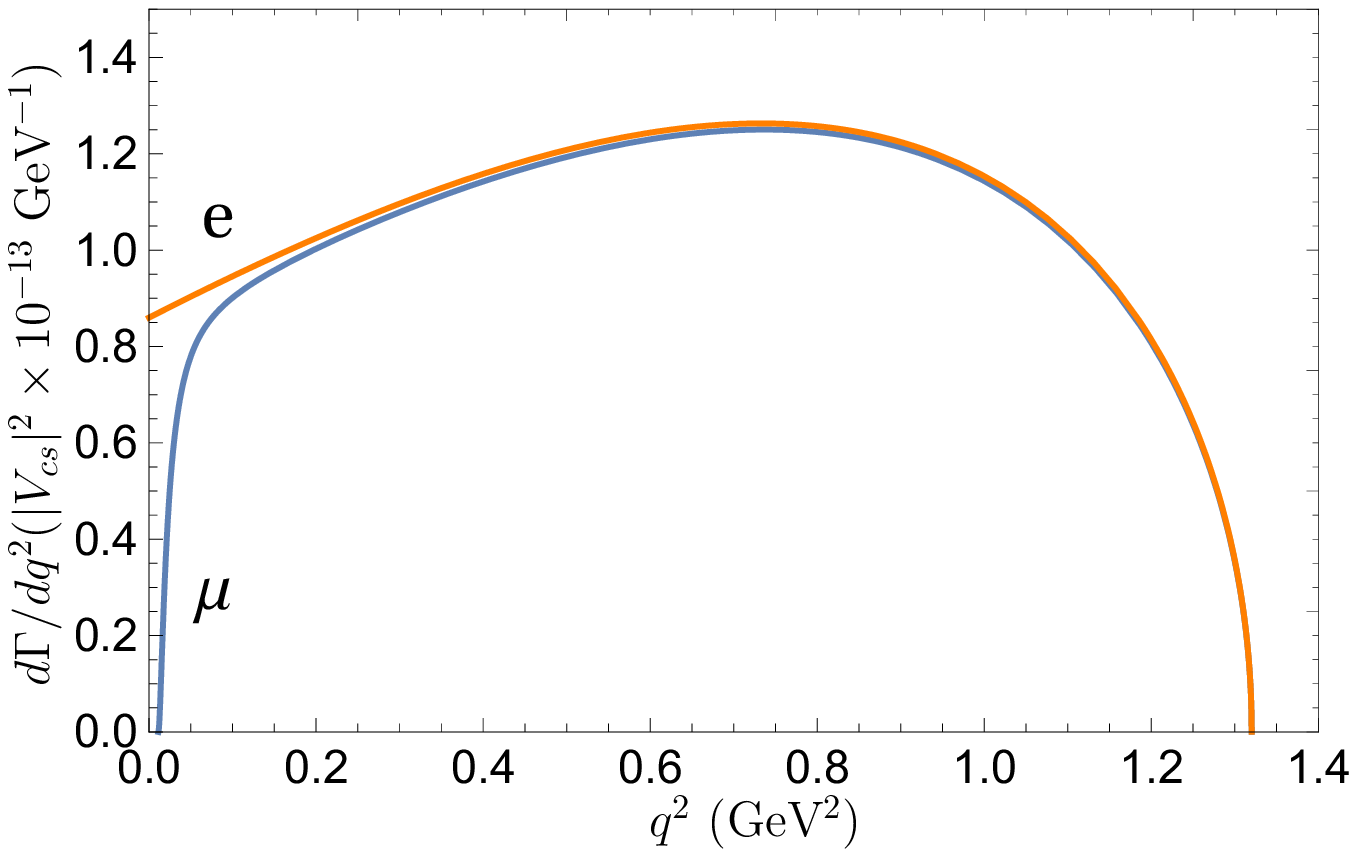}\ \
 \  \includegraphics[width=8cm]{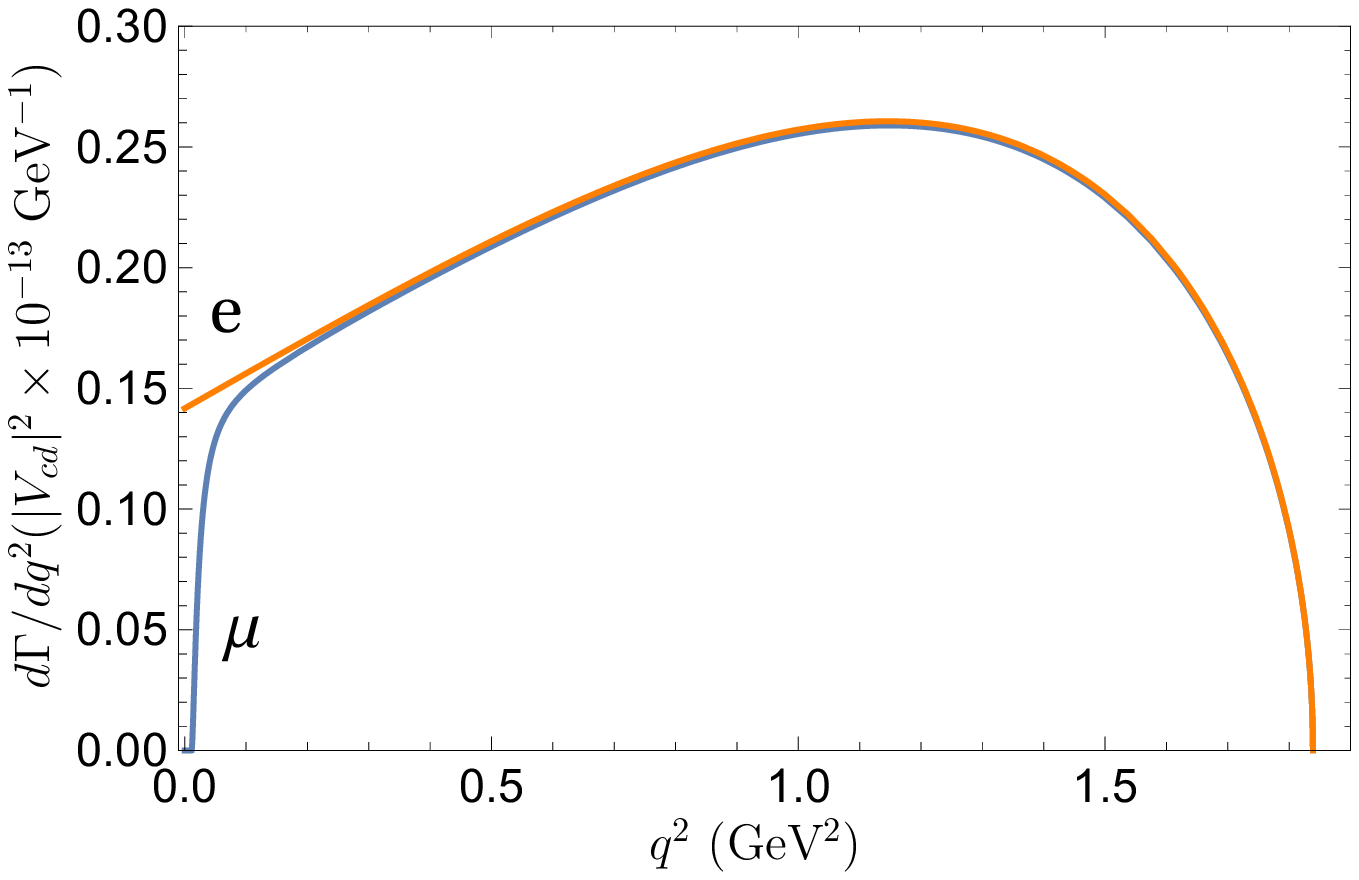}

  \caption{Differential decay rates  of the
    $\Xi_c\to \Xi\ell\nu_\ell$ (left) and $\Xi_c\to \Lambda\ell\nu_\ell$ (right)
    semileptonic decays. }
  \label{fig:brXic}
\end{figure}

The differential and total semileptonic decay rates, branching
fractions and  different asymmetry and polarization parameters can be calculated using
the decay form factors and
the helicity formalism \cite{giklsh}. The relation between helicity
amplitudes and decay form factors are the following.
\begin{eqnarray}
  \label{eq:haa}
 H_{+\frac12 0}^{V,A}&=&\frac{\sqrt{(M_{\Xi_c}\mp M_{\Xi(\Lambda)})^2-q^2}}{\sqrt{q^2}}
  \bigg[(M_{\Xi_c}\pm M_{\Xi(\Lambda)}) f_1^{V,A}(q^2)\pm \frac{q^2}{M_{\Xi_c}} f_2^{V,A}(q^2)\bigg],
\cr
H_{+\frac12 +1}^{V,A}&=&\sqrt{2[(M_{\Xi_c}\mp M_{\Xi(\Lambda)})^2-q^2]}
  \bigg[f_1^{V,A}(q^2)\pm \frac{M_{\Xi_c}\pm M_{\Xi(\Lambda)}}{M_{\Xi_c}}f_2^{V,A}(q^2)\bigg],
\cr
 H_{+\frac12 t}^{V,A}&=&\frac{\sqrt{(M_{\Xi_c}\pm M_{\Xi(\Lambda)})^2-q^2}}{\sqrt{q^2}}
  \bigg[ (M_{\Xi_c}\mp M_{\Xi(\Lambda)}) f_1^{V,A}(q^2)\pm \frac{q^2}{M_{\Xi_c}} f_3^{V,A}(q^2)\bigg],
\end{eqnarray}
and the amplitudes for negative values of the helicities are obtained
from the relations
$$H^{V,A}_{-\lambda',\,-\lambda_W}=\pm H^{V,A}_{\lambda',\, \lambda_W}.$$
The total helicity amplitude for the
$V-A$ current is given by
\begin{equation}
\label{ha}
H_{\lambda',\, \lambda_W}=H^{V}_{\lambda',\, \lambda_W}
-H^{A}_{\lambda',\, \lambda_W}.
\end{equation}
The differential decay rates and angular distributions are expressed in
terms of the  helicity structures which are the following combinations
of the total helicity amplitudes (\ref{ha})  
\begin{eqnarray}
  \label{eq:hhs}
  {\cal H}_U(q^2)&=&|H_{+1/2,+1}|^2+|H_{-1/2,-1}|^2,\cr
{\cal H}_L(q^2)&=&|H_{+1/2,0}|^2+|H_{-1/2,0}|^2,\cr
{\cal H}_S(q^2)&=&|H_{+1/2,t}|^2+|H_{-1/2,t}|^2,\cr
{\cal H}_{SL}(q^2)&=&{\rm Re}(H_{+1/2,0}H_{+1/2,t}^\dag+H_{-1/2,0}H_{-1/2,t}^\dag),\cr
  {\cal H}_P(q^2)&=&|H_{+1/2,+1}|^2-|H_{-1/2,-1}|^2,\cr
{\cal H}_{L_P}(q^2)&=&|H_{+1/2,0}|^2-|H_{-1/2,0}|^2,\cr
{\cal H}_{S_P}(q^2)&=&|H_{+1/2,t}|^2-|H_{-1/2,t}|^2.
\end{eqnarray}
The expression for the differential decay rate is given by \cite{giklsh}
\begin{equation}
  \label{eq:dgamma}
  \frac{d\Gamma(\Xi_c\to \Xi(\Lambda)\ell\bar\nu_\ell)}{dq^2}=\frac{G_F^2}{(2\pi)^3}
  |V_{cq}|^2
  \frac{\lambda^{1/2}(q^2-m_\ell^2)^2}{48M_{\Xi_c}^3q^2}{\cal H}_{tot}(q^2),
\end{equation}
where $G_F$ is the Fermi constant, $V_{cq}$ is the Cabibbo-Kobayashi-Maskawa (CKM) matrix element ($q=s,d$),
 $\lambda\equiv
\lambda(M_{\Xi_c}^2,M_{\Xi(\Lambda)}^2,q^2)=M_{\Xi_c}^4+M_{\Xi(\Lambda)}^4+q^4-2(M_{\Xi_c}^2M_{\Xi(\Lambda)}^2+M_{\Xi(\Lambda)}^2q^2+M_{\Xi_c}^2q^2)$,
and $m_\ell$ is the lepton mass ($\ell=e,\mu$),
\begin{equation}
 \label{eq:hh}
 {\cal H}_{tot}(q^2)=[{\cal H}_U(q^2)+{\cal H}_L(q^2)] \left(1+\frac{m_\ell^2}{2q^2}\right)+\frac{3m_\ell^2}{2q^2}{\cal H}_S(q^2) .
\end{equation}
Substituting in these expressions the $\Xi_c$ decay form factors
calculated in the previous section we obtain the differential decay
rates. We plot them for the $\Xi_c\to
\Xi\ell\nu_\ell$ (left) and $\Xi_c\to \Lambda\ell\nu_\ell$ (right)
semileptonic decays  in Fig.~\ref{fig:brXic}.

\begin{figure}[hbt]
  \centering
 \includegraphics[width=8cm]{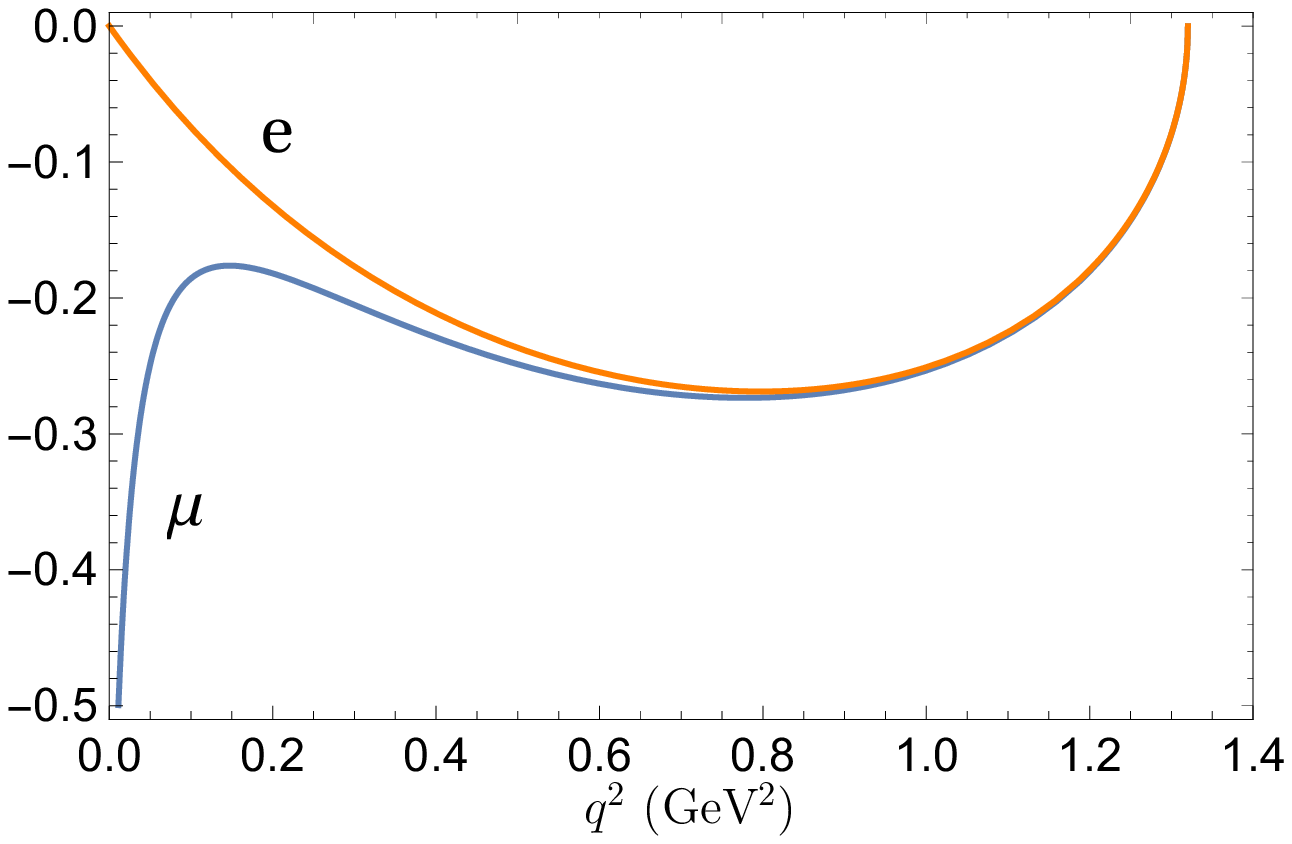}\ \
 \  \includegraphics[width=8cm]{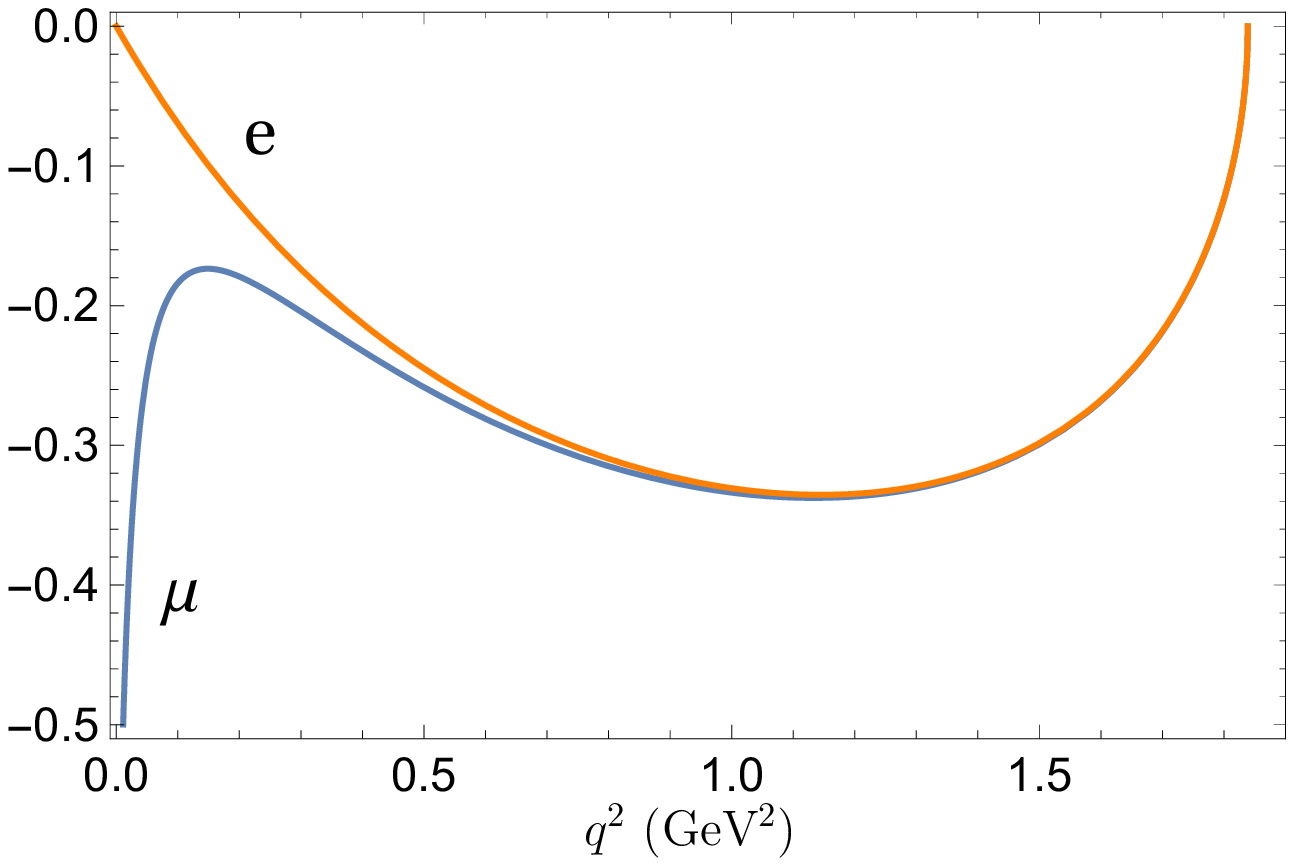}

  \caption{Forward-backward asymmetry $A_{FB}(q^2)$ in the
    $\Xi_c\to \Xi\ell^+\nu_\ell$ (left) and $\Xi_c\to \Lambda\ell^+\nu_\ell$ (right)
    semileptonic decays. }
  \label{fig:afbXib}
\end{figure}

\begin{figure}[hbt]
  \centering
 \includegraphics[width=8cm]{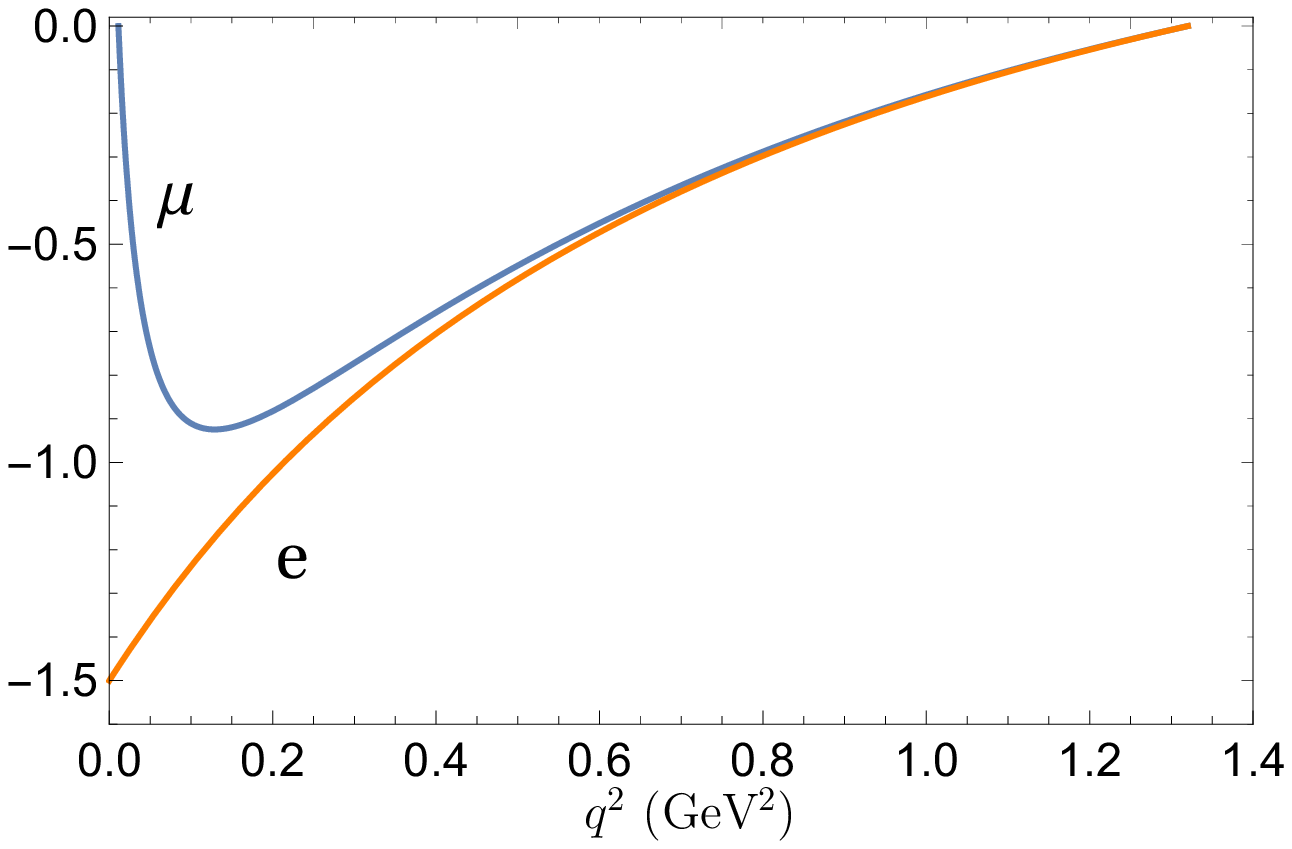}\ \
 \  \includegraphics[width=8cm]{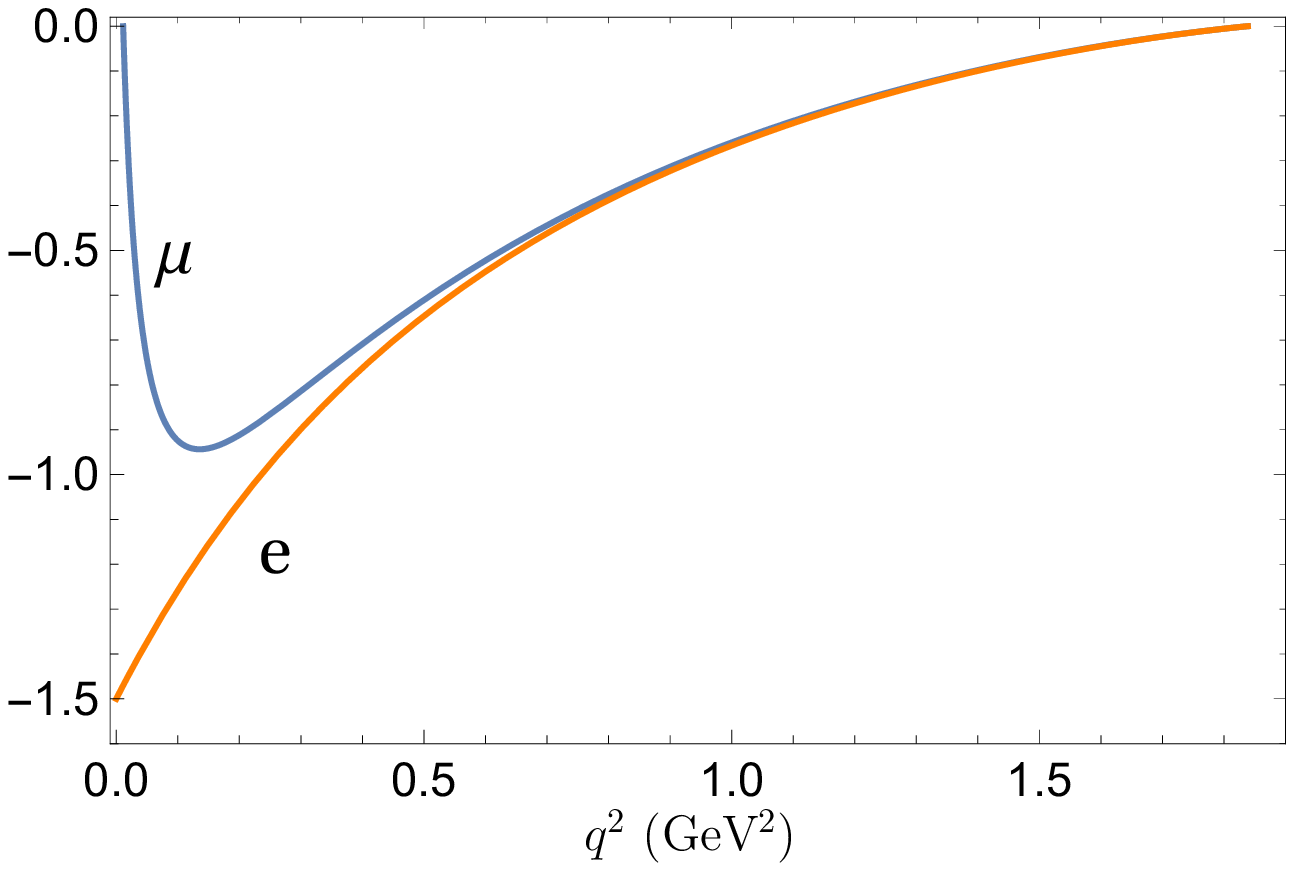}

  \caption{Convexity parameter  $C_F(q^2)$ in the
    $\Xi_c\to \Xi\ell^+\nu_\ell$ (left) and $\Xi\to \Lambda\ell^+\nu_\ell$ (right)
    semileptonic decays. }
  \label{fig:cfXib}
\end{figure}

\begin{figure}[hbt]
  \centering
 \includegraphics[width=8cm]{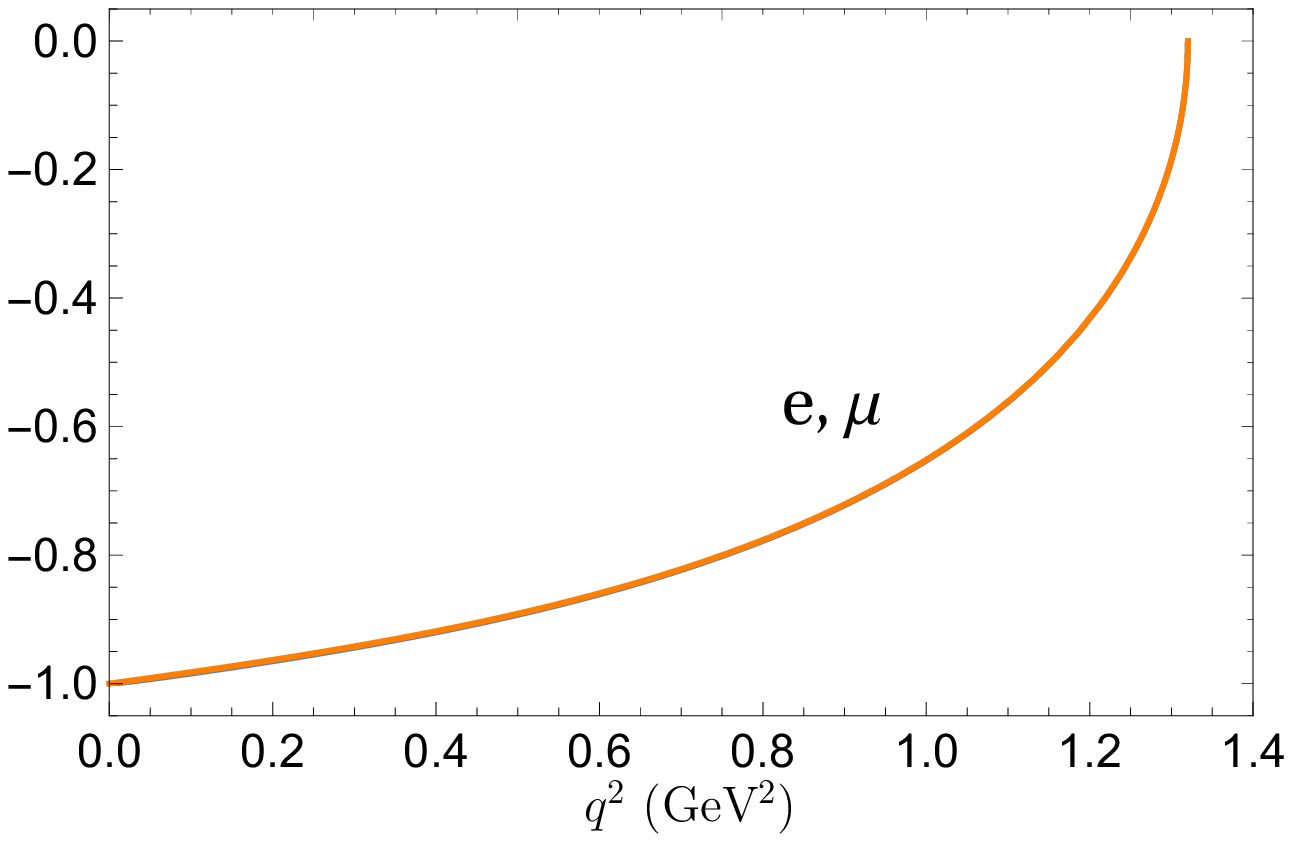}\ \
 \  \includegraphics[width=8cm]{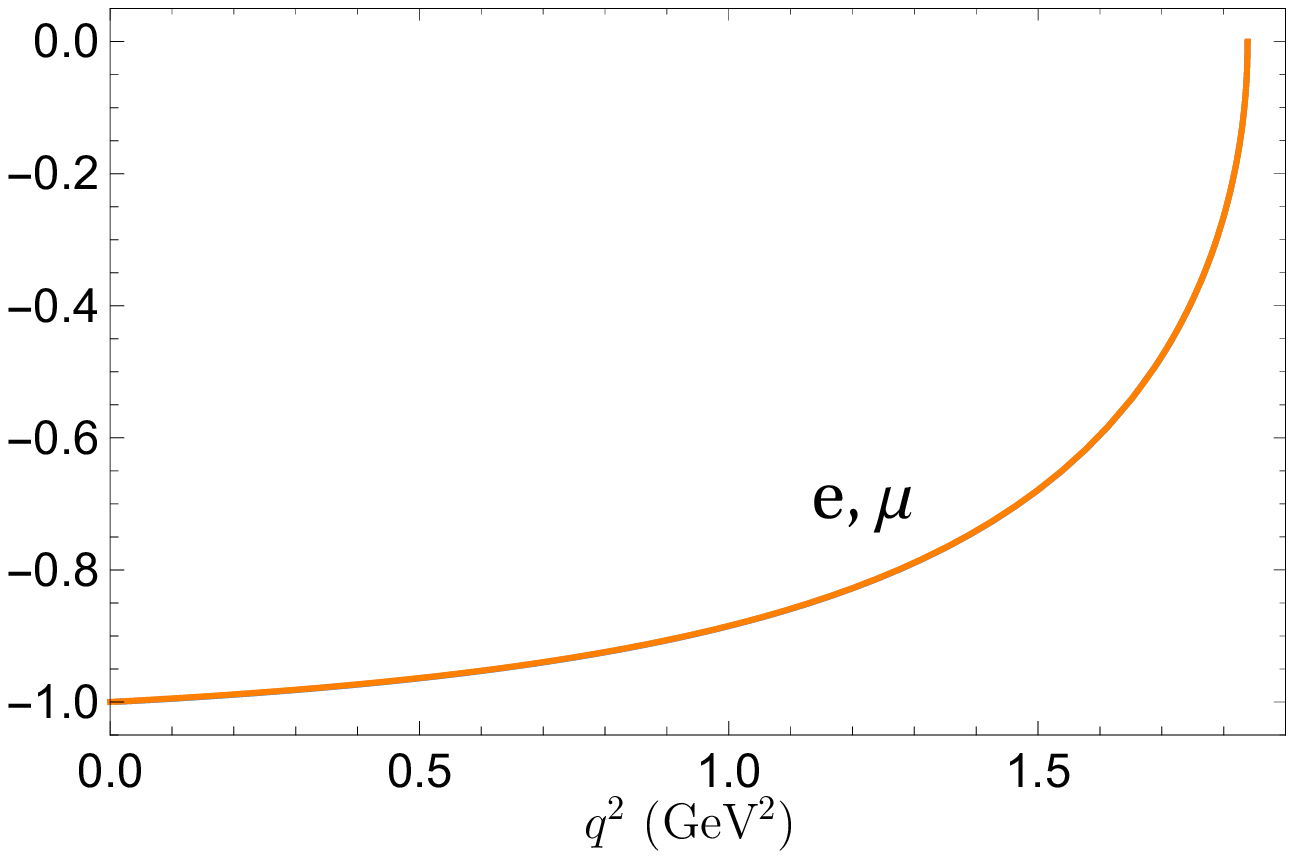}

  \caption{Longitudinal polarization $P_L(q^2)$ of the final baryon  in the
    $\Xi_c\to \Xi\ell^+\nu_\ell$ (left) and $\Xi_c\to \Lambda\ell^+\nu_\ell$ (right)
    semileptonic decays. }
  \label{fig:PLXib}
\end{figure}

\begin{figure}[hbt]
  \centering
 \includegraphics[width=8cm]{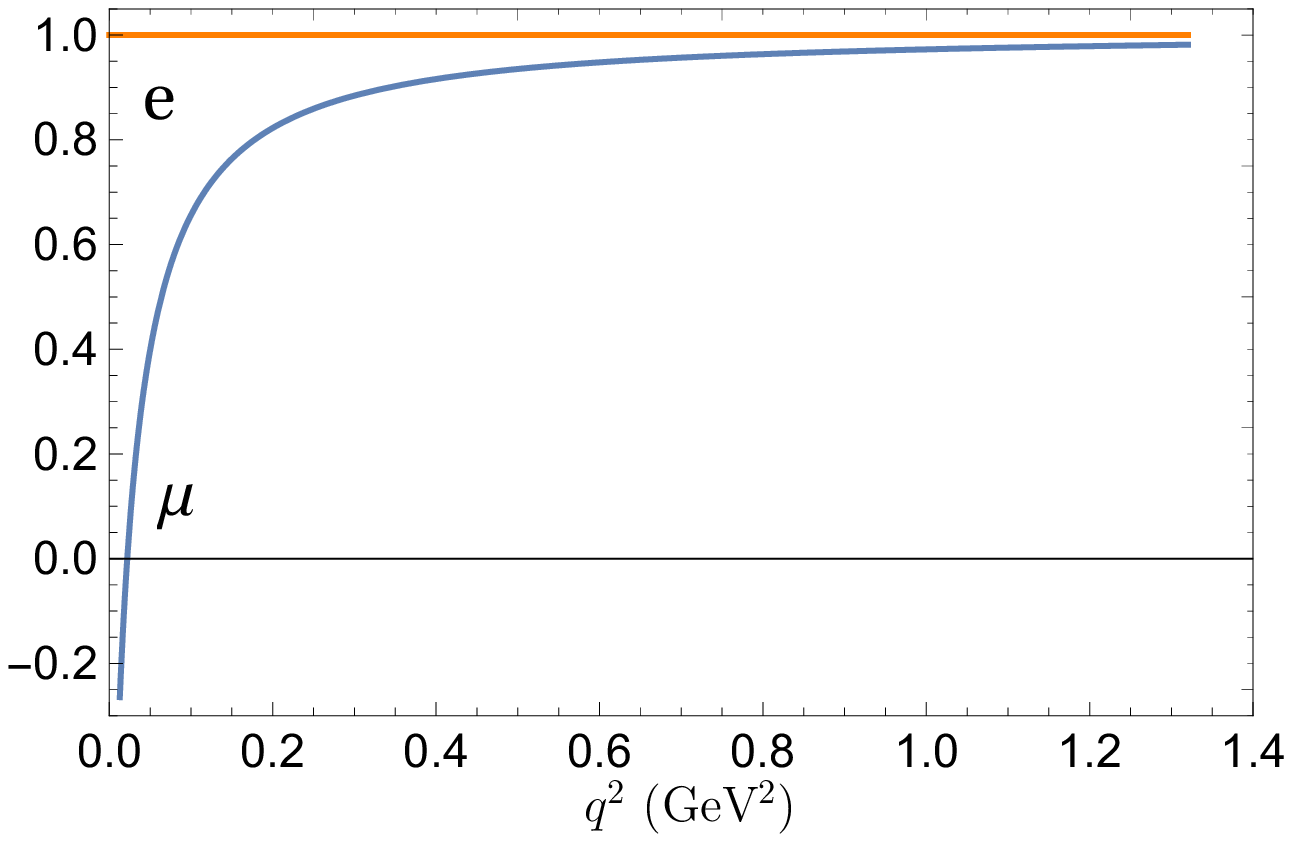}\ \
 \  \includegraphics[width=8cm]{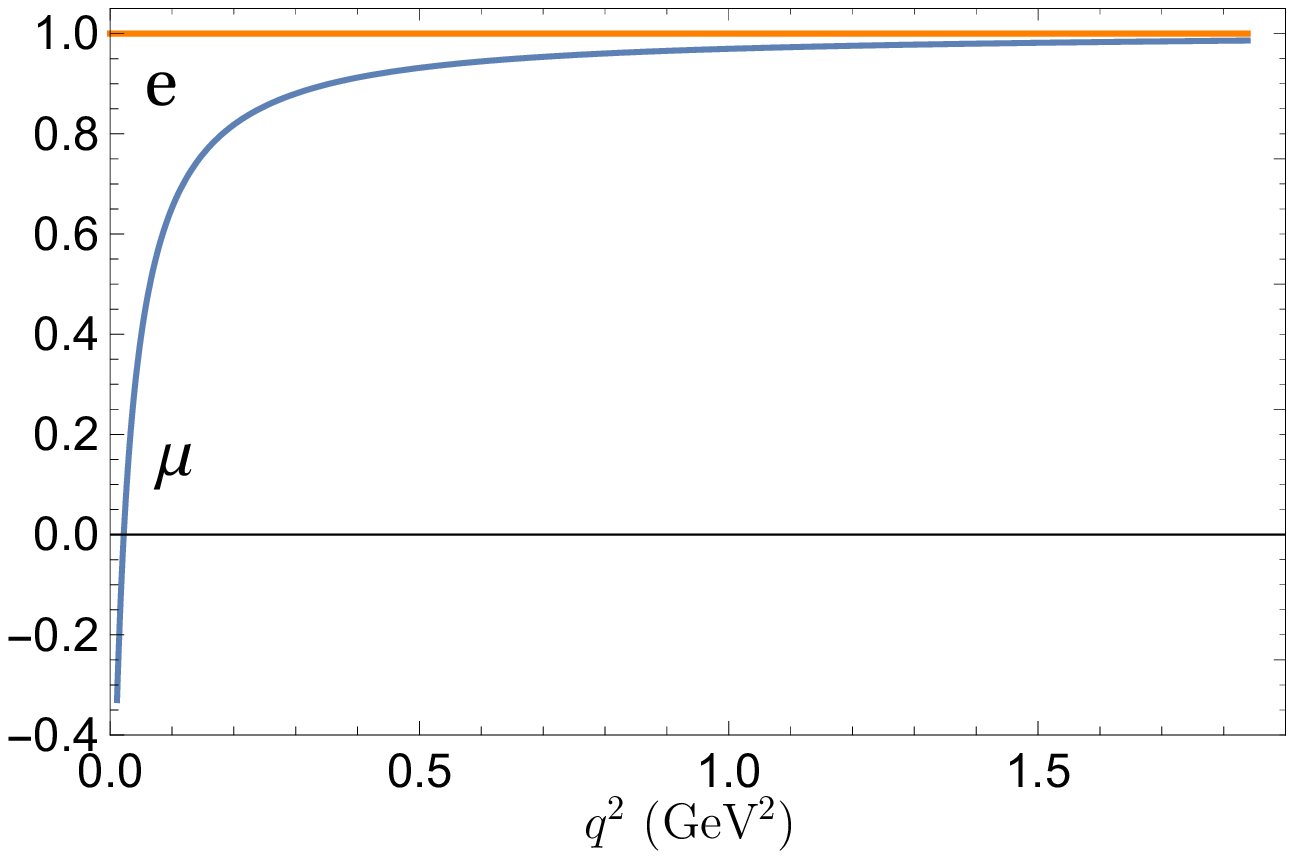}

  \caption{Longitudinal polarization $P_\ell(q^2)$
    of the charged lepton  in the
    $\Xi_c\to \Xi\ell^+\nu_\ell$ (left) and $\Xi_c\to \Lambda\ell^+\nu_\ell$ (right)
    semileptonic decays. }
  \label{fig:PellXib}
\end{figure}

Many important observables are expressed in terms of the
helicity combinations (\ref{eq:hhs}) (see \cite{giklsh} for details):\\
\begin{itemize}
\item The forward-backward asymmetry of the charged lepton 
\begin{equation}
  \label{eq:afb}
  A_{FB}(q^2)=\frac{\frac{d\Gamma}{dq^2}({\rm forward})-\frac{d\Gamma}{dq^2}({\rm backward})}{\frac{d\Gamma}{dq^2}}
=\frac34\frac{{\cal H}_P(q^2)-2\frac{m_\ell^2}{q^2}{\cal H}_{SL}(q^2)}{{\cal H}_{tot}(q^2)};\qquad
\end{equation}
\item The convexity parameter  
\begin{equation}
  \label{eq:cf}
  C_F(q^2)=\frac34\left(1-\frac{m_\ell^2}{q^2}\right)\frac{{\cal H}_U(q^2)-2{\cal H}_L(q^2)}{{\cal H}_{tot}(q^2)};
\end{equation}
\item The longitudinal polarization of the final baryon $\Xi_c(\Lambda)$ 
  \begin{equation}
    \label{eq:pl}
P_L(q^2)=\frac{[{\cal H}_P(q^2)+{\cal H}_{L_P}(q^2)]\left(1+\frac{m_\ell^2}{2q^2}\right)+3 \frac{m_\ell^2}{2q^2}{\cal H}_{S_P}(q^2)}{{\cal H}_{tot}(q^2)};
\end{equation}
\item The longitudinal polarization of the charged lepton $\ell$ 
  \begin{equation}
     \label{eq:pel}
P_\ell(q^2)=\frac{{\cal H}_U(q^2)+{\cal H}_L(q^2)-\frac{m_\ell^2}{2q^2}[{\cal H}_U(q^2)+{\cal H}_L(q^2)+3{\cal H}_{S}(q^2)]}{{\cal H}_{tot}(q^2)}.
\end{equation}
\end{itemize}
These observables are plotted for $\Xi_c\to\Xi\ell^+\nu_\ell$ and 
$\Xi_c\to \Lambda\ell^+\nu_\ell$ semileptonic decays in Figs.~\ref{fig:afbXib}-\ref{fig:PellXib}. The predictions
for the decay branching fractions and asymmetry parameters are
presented in Table~\ref{drbr}.  We calculated the decay rates  using the CKM values
$|V_{cs}|=0.995\pm0.016$, 
$|V_{cd}|=0.220\pm 0.005$  \cite{pdg}. The average values of the
$\langle
A_{FB}\rangle$,  $\langle C_F\rangle$, $\langle P_L\rangle$ and
 $\langle P_\ell\rangle$
were obtained by separately integrating the numerators and
denominators in Eqs.~(\ref{eq:afb})-(\ref{eq:pel}) over $q^2$. We
roughly estimate uncertainties of our predictions for the branching
fractions to be about 10\%.

\begin{table}
\caption{$\Xi_c$ semileptonic branching fractions,
  asymmetry and polarization parameters. }
\label{drbr}
\begin{ruledtabular}
\begin{tabular}{cccccc}
Decay&  $Br$ (\%)
  & $\langle A_{FB}\rangle$ &$\langle C_F\rangle$& $\langle P_L\rangle$& $\langle P_\ell\rangle$\\
\hline
$\Xi_c^0\to\Xi^-e^+\nu_e$ & 2.38 & $-0.208$&$-0.519$&$-0.795$&$1$\\
$\Xi_c^0\to\Xi^-\mu^+\nu_\mu$  & 2.31
  &$-0.235$&$-0.436$&$-0.791$&$0.909$\\
  $\Xi_c^+\to\Xi^0e^+\nu_e$ & 9.40 & $-0.208$&$-0.519$&$-0.795$&$1$\\
$\Xi_c^+\to\Xi^0\mu^+\nu_\mu$  & 9.11  &$-0.235$&$-0.436$&$-0.791$&$0.909$\\
$\Xi_c^+\to \Lambda e^+\nu_e$  & 0.127  & $-0.266$&$-0.441$&$-0.842$&$1$\\
$\Xi_c^+\to \Lambda\mu^+\nu_\mu$ & 0.124  & $-0.283$&$-0.384$&$-0.841$&$0.935$\\
\end{tabular}
\end{ruledtabular}
\end{table}

Since possible violations of the charged lepton universality are now
widely discussed in the literature \cite{bifani} we present predictions for the
corresponding ratios for the semileptonic $\Xi_c$ decays involving muons and electrons 
\begin{eqnarray}
  \label{eq:ratio}
  R_{\Xi}&=&\frac{Br(\Xi_c\to\Xi\mu\nu_\mu)}{Br(\Xi_c\to\Xi
  e\nu_e)}=0.969\pm 0.010,\cr
R_{\Lambda}&=&\frac{Br(\Xi_c\to \Lambda\mu\nu_\mu)}{Br(\Xi_c\to \Lambda
               e\nu_e)}=0.980\pm0.010.
\end{eqnarray}
Most of the theoretical uncertainties cancels in these ratios, thus we
roughly estimate them to be about 1\%.

Theoretical predictions \cite{zhao,ghlt,glty,ass} for the semileptonic
$\Xi_c$ decay branching fractions are compared in
Table~\ref{comp}. The light-front quark model is used for the weak decay form
factor and branching fraction calculations in Ref.~\cite{zhao}. The
predictions of Refs.~ \cite{ghlt,glty} are based on the application of
the SU(3) flavor symmetry, while the light-cone QCD sum rules are
employed in Ref.~\cite{ass}. The experimental branching fractions are
obtained by multiplying the CLEO~II values \cite{cleo} updated by PDG \cite{pdg} for the ratios
$\Gamma(\Xi_c^0\to \Xi^-e^+\nu_e)/\Gamma(\Xi_c^0\to
\Xi^-\pi^+)=3.1\pm1.1$ and $\Gamma(\Xi_c^+\to \Xi^0e^+\nu_e)/\Gamma(\Xi_c^+\to
\Xi^-\pi^+\pi^+)=2.3^{+0.7}_{-0.8}$ by the recently measured by the Belle
Collaboration branching fractions $Br(\Xi_c^0\to
\Xi^-\pi^+)=(1.80\pm0.50\pm01.14)\%$ \cite{belle1} and $Br(\Xi_c^+\to
\Xi^-\pi^+\pi^+)=(2.86\pm1.21\pm0.38)\%$ \cite{belle2}. We find reasonable agreement
of our predictions for the CKM favored $\Xi_c\to\Xi \ell\nu_\ell$ decays with the
results of Refs.~\cite{zhao,ghlt,glty} and experimental data, while
the light-cone QCD sum rule values for branching fractions are substantially higher and
disagree with data by more than a factor of 2  for the
$\Xi_c^+\to\Xi^0e^+\nu_e$ decay. Note that ARGUS
Collaboration \cite{argus} measured  the
ratio $\Gamma(\Xi_c^0\to \Xi^-e^+{\rm anything})/\Gamma(\Xi_c^0\to
\Xi^-\pi^+)=1.0\pm0.5$ (the value updated by PDG \cite{pdg}) which
combined with Belle data \cite{belle1} leads to the semi-inclusive
branching fraction $Br(\Xi_c^0\to \Xi^-e^+{\rm
  anything})=(1.80\pm1.07)\%$ in  good agreement with our
result. There is no data yet for the CKM suppressed $\Xi_c^+\to
\Lambda \ell^+\nu_\ell$ decays and theoretical evaluations give
consistent predictions for the branching fractions about 0.1\%.

\begin{table}
\caption{Comparison of theoretical predictions for the
  $\Xi_c$ semileptonic decay  branching fractions (in \%) with available
  experimental data.}
\label{comp}
\begin{ruledtabular}
\begin{tabular}{ccccccc}
Decay& this paper&\cite{zhao}&\cite{ghlt} & \cite{glty}&\cite{ass}&Experiment\\
\hline
$Br(\Xi_c^0\to\Xi^-e^+\nu_e)$ & 2.38& 1.35&$4.87\pm1.74$&$2.4\pm0.3$&$7.26\pm2.54$ & $5.58\pm2.62$\\
$Br(\Xi_c^0\to\Xi^-\mu^+\nu_\mu)$ &2.31& & &$2.4\pm0.3$&$7.15\pm2.50$ \\
$Br(\Xi_c^+\to\Xi^0e^+\nu_e)$ & 9.40& 5.39&$3.38^{+2.19}_{-2.26}$&$9.8\pm1.1$ &$28.6\pm10.0$ &$6.58\pm3.85$\\
  $Br(\Xi_c^+\to\Xi^0\mu^+\nu_\mu)$ &9.11& & &$9.8\pm1.1$&$28.2\pm9.9$\\
  $Br(\Xi_c^+\to \Lambda e^+\nu_e)$  & 0.127& 0.082& &$0.166\pm0.018$\\
  $Br(\Xi_c^+\to \Lambda \mu^+\nu_\mu)$  & 0.124&\\
\end{tabular}
\end{ruledtabular}
\end{table}

We can combine our present predictions for the semileptonic $\Xi_c$
baryon decays with our previous analysis of the semileptonic
$\Lambda_c$ decays \cite{Lambdacsl} to test the flavor $SU(3)$
symmetry. Under the exact $SU(3)$ limit the following relations should hold
 \cite{zhao,ghlt}
\begin{equation}
  \label{eq:su3}
  \frac{\Gamma(\Lambda_c\to n e\nu_e)}{|V_{cd}|^2}=\frac{3\Gamma(\Lambda_c\to \Lambda e\nu_e)}{2|V_{cs}|^2}=\frac{6\Gamma(\Xi_c\to \Lambda e\nu_e)}{|V_{cd}|^2}=\frac{\Gamma(\Xi_c\to \Xi e\nu_e)}{|V_{cs}|^2}.
\end{equation}
In Table~\ref{su3} we test these relations for the ratios
$\Gamma/|V_{cq}|^2$, where the corresponding CKM matrix element is
used. In first column we give the semileptonic decay process. In the
second column we present predictions of our model. The third column
contains the flavor $SU(3)$ symmetry result using relations
(\ref{eq:su3}) and $\Gamma(\Lambda_c\to n e\nu_e)/|V_{cd}|^2$ as an
input, while the last column gives the relative difference in \%. From
this table we see that the flavor $SU(3)$ is broken for the charmed
baryon semileptonic decays especially for $\Xi_c$ where its breaking
is about 20-35\%. This is the consequence of the larger mass of the
$s$ quark in comparison with the $u,d$ quarks and the employed quark-diquark
picture of baryons.       

\begin{table}
\caption{Predictions for the ratios $\Gamma/|V_{cq}|^2$ in ps$^{-1}$ ($q=s,d$)}
\label{su3}
\begin{ruledtabular}
\begin{tabular}{cccc}
  Decay&our result& exact $SU(3)$&difference\\
  \hline
  $\Lambda_c\to n e\nu_e$& 0.265& 0.265& \\
  $\Lambda_c\to \Lambda e\nu_e$& 0.167 & 0.177& 6\%\\
  $\Xi_c\to \Lambda e\nu_e$ & 0.059& 0.044& 34\%\\
  $\Xi_c\to \Xi e\nu_e$& 0.215 & 0.265& 19\%
\end{tabular}
\end{ruledtabular}
\end{table}

\section{Conclusion}

The relativistic quark model was used for the calculation of  form
factors of the
semileptonic $\Xi_c$ transitions, both for the CKM favored
$\Xi_c\to\Xi \ell^+\nu_\ell$ and CKM suppressed $\Xi_c^+\to
\Lambda \ell^+\nu_\ell$ decays. All relativistic effects, including
baryon wave function transformations from the rest to moving reference
frame and contributions of the intermediate negative-energy states,
were comprehensively taken into account. This allowed us to  explicitly
determine the momentum transfer $q^2$ dependence of the weak form
factors in the whole kinematical range without additional model
assumptions or extrapolations. We give the values of parameters in the
analytic expression (\ref{fitff}) which accurately approximates the numerically calculated 
form factors in Tables~\ref{ffXicXi}, \ref{ffXicLambda} and show
the form factor $q^2$ dependence in Figs.~\ref{fig:ffXicXi},
\ref{fig:ffXicLambda}.

Using the calculated form factors and helicity formalism we estimated
important observables for the CKM favored and CKM suppressed  $\Xi_c$
semileptonic decays: differential decay rates, branching fractions, different asymmetry and
polarization parameters, which are given in Table~\ref{drbr} and
plotted in Figs.~\ref{fig:afbXib}-\ref{fig:PellXib}. Our results for
the branching fractions of the  $\Xi_c\to\Xi \ell\nu_\ell$ decays are
in reasonable agreement with previous calculations based on the
light-front quark model \cite{zhao} and application of the SU(3)
flavor symmetry \cite{ghlt,glty}, but significantly lower than the
light-cone QCD sum rule predictions \cite{ass}. They agree  with
experimental values which can be obtained combining the corresponding
decay ratios from PDG \cite{pdg} and recent measurements of absolute branching
fractions by the Belle Collaboration \cite{belle1,belle2}. Comparison
of the present results for the semileptonic $\Xi_c$ decays with the
previous ones for the $\Lambda_c$ decays \cite{Lambdacsl} indicates a
sizable $SU(3)$ symmetry breaking which can reach 35\%. This
result is in accord with the conclusion of Ref.~\cite{zhao}.

In this paper we limit our consideration to the $\Xi_c$ semileptonic
decays. In principle it can be extended to $\Xi_c'$ baryons which
differ by the spin of the diquark. However $\Xi_c'$ can decay
radiatively to $\Xi_c$ with the evaluated width of a few keV (see
e.g. \cite{wmz} and references therein). A rough estimate of the
$\Xi_c'$ semileptonic decay rates based on the heavy quark and flavor
$SU(3)$ symmetries indicate that they should be of the same order as
the   $\Xi_c$ semileptonic decay rates. Note that similar results were obtained
in the light cone QCD sum rules \cite{ass}. Therefore the branching
fractions of the $\Xi_c'$ semileptonic decays are expected to be of
order of $10^{-8}-10^{-7}$. Taking into account that the total width of
the $\Xi_c'$ baryon is not measured yet \cite{pdg} due to the poor
statistics it is very unlikely that decays with such small branching
fractions will be observed experimentally.

\acknowledgments
The authors are grateful to D. Ebert and M. Ivanov  for valuable  discussions.


\begin{thebibliography}{00}
\bibitem{pdg} 
  M.~Tanabashi {\it et al.} [Particle Data Group],
  ``Review of Particle Physics,''
  Phys.\ Rev.\ D {\bf 98}, no. 3, 030001 (2018).
\bibitem{belle1} 
  Y.~B.~Li {\it et al.} [Belle Collaboration],
  ``First Measurements of Absolute Branching Fractions of the $\Xi_c^0$ Baryon at Belle,''
  Phys.\ Rev.\ Lett.\  {\bf 122}, no. 8, 082001 (2019).
\bibitem{belle2} 
  Y.~B.~Li {\it et al.} [Belle Collaboration],
  ``First measurements of absolute branching fractions of the $\Xi_c^+$ baryon at Belle,''
  arXiv:1904.12093 [hep-ex].
\bibitem{Lambdabsl} 
  R.~N.~Faustov and V.~O.~Galkin,
  ``Semileptonic decays of $\Lambda_b$ baryons in the relativistic quark model,''
  Phys.\ Rev.\ D {\bf 94}, no. 7, 073008 (2016).
  \bibitem{xib} 
  R.~N.~Faustov and V.~O.~Galkin,
  ``Relativistic description of the $\Xi_b$ baryon semileptonic decays,''
  Phys.\ Rev.\ D {\bf 98}, no. 9, 093006 (2018).
\bibitem{Lambdacsl} 
  R.~N.~Faustov and V.~O.~Galkin,
  ``Semileptonic decays of $\Lambda _c$ baryons in the relativistic quark model,''
  Eur.\ Phys.\ J.\ C {\bf 76}, no. 11, 628 (2016).

  \bibitem{giklsh} 
T.~Gutsche, M.~A.~Ivanov, J.~G.~K\"orner, V.~E.~Lyubovitskij, P.~Santorelli and N.~Habyl,
  ``Semileptonic decay $\Lambda_b \to \Lambda_c + \tau^- + \bar{\nu_\tau}$ in the covariant confined quark model,''
  Phys.\ Rev.\ D {\bf 91}, no. 7, 074001 (2015)
  Erratum: [Phys.\ Rev.\ D {\bf 91}, no. 11, 119907 (2015)].

  \bibitem{hbarregge} 
  D.~Ebert, R.~N.~Faustov and V.~O.~Galkin,
  ``Spectroscopy and Regge trajectories of heavy baryons in the relativistic quark-diquark picture,''
  Phys.\ Rev.\ D {\bf 84}, 014025 (2011).
\bibitem{sbar}
  R.~N.~Faustov and V.~O.~Galkin,
  ``Strange baryon spectroscopy in the relativistic quark model,''
  Phys.\ Rev.\ D {\bf 92}, no. 5, 054005 (2015).
\bibitem{zhao} 
  Z.~X.~Zhao,
  ``Weak decays of heavy baryons in the light-front approach,''
  Chin.\ Phys.\ C {\bf 42}, no. 9, 093101 (2018).  
\bibitem{ass} 
  K.~Azizi, Y.~Sarac and H.~Sundu,
  ``Light cone QCD sum rules study of the semileptonic heavy $\Xi_{Q}$ and $\Xi'_{Q}$ transitions to $\Xi$ and $\Sigma $ baryons,''
  Eur.\ Phys.\ J.\ A {\bf 48}, 2 (2012).

\bibitem{bifani} 
  S.~Bifani, S.~Descotes-Genon, A.~Romero Vidal and M.~H.~Schune,
  ``Review of Lepton Universality tests in $B$ decays,''
  J.\ Phys.\ G {\bf 46}, no. 2, 023001 (2019).
\bibitem{ghlt} 
  C.~Q.~Geng, Y.~K.~Hsiao, C.~W.~Liu and T.~H.~Tsai,
  ``Antitriplet charmed baryon decays with SU(3) flavor symmetry,''
  Phys.\ Rev.\ D {\bf 97}, no. 7, 073006 (2018).
\bibitem{glty} 
  C.~Q.~Geng, C.~W.~Liu, T.~H.~Tsai and S.~W.~Yeh,
  ``Semileptonic decays of anti-triplet charmed baryons,''
  Phys.\ Lett.\ B {\bf 792}, 214 (2019).

\bibitem{cleo} 
  J.~P.~Alexander {\it et al.} [CLEO Collaboration],
  ``First observation of $\Xi_c^+ \to \Xi^0 e^+ \nu_e$ and a measurement of the $\Xi_c^+ / \Xi_c^0$ lifetime ratio,''
  Phys.\ Rev.\ Lett.\  {\bf 74}, 3113 (1995)
  Erratum: [Phys.\ Rev.\ Lett.\  {\bf 75}, 4155 (1995)].
\bibitem{argus} 
  H.~Albrecht {\it et al.} [ARGUS Collaboration],
  ``Observation of $\Xi_c^0$ semileptonic decay,''
  Phys.\ Lett.\ B {\bf 303}, 368 (1993).

\bibitem{wmz} 
  G.~J.~Wang, L.~Meng and S.~L.~Zhu,
  ``Radiative decays of the singly heavy baryons in chiral perturbation theory,''
  Phys.\ Rev.\ D {\bf 99}, no. 3, 034021 (2019).
  

 
  
\end{thebibliography}
\end{document}